\documentclass[conference,letterpaper]{IEEEtran}

\usepackage{algorithm}
\usepackage{algpseudocode}
\usepackage[utf8]{inputenc} 
\usepackage[T1]{fontenc}
\usepackage{url}
\usepackage{ifthen}
\usepackage{graphicx}
\usepackage{cite}
\usepackage{color}
\usepackage[cmex10]{amsmath} 
\usepackage{amssymb,amsfonts,amsthm}

\newtheorem{theorem}{Theorem}

\newtheorem{proposition}{Proposition}
\newtheorem{lemma}{Lemma}
\newtheorem{corollary}{Corollary}

\begin{document}

\title{Approaching Maximum Likelihood Decoding Performance via Reshuffling ORBGRAND}

\author{%
  \IEEEauthorblockN{Li Wan and Wenyi Zhang}
  \IEEEauthorblockA{University of Science and Technology of China\\
                    Email: wenyizha@ustc.edu.cn}
}


\maketitle


\begin{abstract}
  Guessing random additive noise decoding (GRAND) is a recently proposed decoding paradigm particularly suitable for codes with short length and high rate. Among its variants, ordered reliability bits GRAND (ORBGRAND) exploits soft information in a simple and effective fashion to schedule its queries, thereby allowing efficient hardware implementation. Compared with maximum likelihood (ML) decoding, however, ORBGRAND still exhibits noticeable performance loss in terms of block error rate (BLER). In order to improve the performance of ORBGRAND while still retaining its amenability to hardware implementation, a new variant of ORBGRAND termed RS-ORBGRAND is proposed, whose basic idea is to reshuffle the queries of ORBGRAND so that the expected number of queries is minimized. Numerical simulations show that RS-ORBGRAND leads to noticeable gains compared with ORBGRAND and its existing variants, and is only 0.1dB away from ML decoding, for BLER as low as $10^{-6}$.
\end{abstract}
\let\thefootnote\relax\footnotetext{This work was supported in part by the National Natural Science Foundation of China under Grant 62231022.}

\section{Introduction}
In recent years, a variety of decoding methods generally referred to as guessing random additive noise decoding (GRAND) have been proposed\cite{duffy2019capacity}\cite{duffy2021guessing}\cite{duffy2022ordered}\cite{solomon2020soft}. These decoding methods are universal in the sense that they are applicable to all block codes, building upon the key idea of conducting a sequence of queries to test whether a queried error pattern can convert a noise-corrupted channel output vector into a codeword. They are particularly suitable for codes with relatively short length and high rate\cite{duffy2019capacity}\cite{yue2023efficient}, meeting the requirements of ultra-reliable low-latency communication (URLLC). 
Among the variants of GRAND, soft GRAND (SGRAND) \cite{solomon2020soft}, which schedules queries based on the exact values of log likelihood ratios (LLRs), is in fact equivalent to maximum likelihood (ML) decoding \cite{liu2022orbgrand}\cite[Ch. 10]{Lin2004ErrorCC} (assuming that the queries exhaust all possible error patterns, in the worst case). It, however, is not amenable to hardware implementation because its queries are generated based on the exact values of LLRs in a sequential fashion. On the other hand, ordered reliability bits GRAND (ORBGRAND) \cite{duffy2022ordered} only needs the relationship of ranking among the magnitudes of LLRs to generate queries. Once the ranking is obtained, the queries can then be implemented in a parallel fashion, thereby facilitating efficient hardware implementation \cite{duffy2022ordered}\cite{galligan2023block}\cite{an2023soft}\cite{abbas2022high}\cite{condo2021high}.

From an information-theoretic perspective, ORBGRAND has been shown to be almost capacity-achieving\cite{liu2022orbgrand}. For finite code lengths, however, ORBGRAND still exhibits a noticeable performance loss compared with ML decoding in terms of block error rate (BLER), especially as signal-to-noise ratio (SNR) increases. Several works have focused on improving ORBGRAND while still retaining its amenability to hardware implementation (i.e., using the ranking among magnitudes of LLRs only), by scheduling queries based on various heuristics \cite{duffy2022ordered}\cite{liu2022orbgrand}\cite{condo2021high}, typically related to the cumulative distribution function (CDF) of the magnitude of LLR. Other enhancement techniques include adopting list decoding \cite{abbas2022list} and combining ad hoc queries based on empirical evidences \cite{condo2022fixed}.

In this work, we propose an improvement of ORBGRAND, termed RS-ORBGRAND, motivated by the insight that, if, ideally, one could reshuffle the queries of ORBGRAND so that they are identical to those of SGRAND, then the performance of the thus reshuffled ORBGRAND would be exactly that of SGRAND and hence be optimal. Such an idealistic situation, of course, is impossible, but it provides a guideline for us, and drives us to study the difference between an arbitrary querying schedule and the querying schedule of SGRAND.

In order to conduct a tractable analysis, we turn to a search problem, wherein a randomly selected codeword is corrupted by additive white Gaussian noise (AWGN) and is observed by a searcher, and the searcher conducts a sequence of queries to recover this particular codeword. Such a problem is an idealized decoding problem, without needing to consider the risk of mistakenly deciding an incorrect codeword. For this search problem, we analyze the expected number of queries, showing that the querying schedule that minimizes the expected number of queries obeys a monotonicity property. Consequently, we propose to reshuffle the querying schedule generated by ORBGRAND so as to satisfy this monotonicity property. Since the reshuffling is offline, this method follows the same decoding process as ORBGRAND and has low complexity. When applying the thus obtained RS-ORBGRAND to decoding, numerical simulations using BCH and polar codes show that it leads to a gain of no less than 0.3dB compared with ORBGRAND and its existing variants, and is only 0.1dB away from ML decoding, for BLER as low as $10^{-6}$.

The remaining part of this paper is organized as follows. Section II introduces the channel model and GRAND. Section III analyzes the search problem motivated by GRAND, and proposes RS-ORBGRAND. Section IV applies RS-ORBGRAND to decoding, and verifies its effectiveness via numerical simulations. Section V concludes this paper. We use uppercase letters (e.g., $X$) to represent random variables and their corresponding lowercase letters (e.g., $x$) to represent their realizations.

\section{Preliminaries}

\subsection{Channel Model}\label{2.1}

We consider general binary block coding wherein information bits $\underline{U} \in \mathbb{F}_{2}^{k}$ generate a codeword $\underline{W} \in \mathcal{C} \subseteq \mathbb{F}_{2}^{n}$. The code rate is thus $R = \frac{k}{n}$. For simplicity, in this work we consider BPSK modulation over an AWGN channel, while the basic idea should be applicable, with appropriate generalizations, to general memoryless channels. We assume that each element of $\underline{W}$, $W_i$, $i = 1, \ldots, n$, takes $0$ and $1$ with equal probability, a condition commonly satisfied for linear block codes. With BPSK, the transmitted vector $\underline{X}$ is such that $X_i = 1$ if $W_i = 0$ and $X_i = -1$ if $W_i = 1$, $i = 1, \ldots, n$; and the resulting AWGN channel output vector satisfies $\underline{Y}|\underline{X} \sim \mathcal{N}(\underline{X}, \sigma^2 \mathbf{I}_{n \times n})$. Given $\underline{Y}$, the LLRs are
\begin{equation*}\label{eq:LLR}
    L_{i} = \log\frac{P(Y_i \mid W_{i} = 0)}{P(Y_i \mid W_{i} = 1)} = \frac{2}{\sigma^2} Y_i, \  i = 1, \ldots, n,
\end{equation*}
and subsequently we use $\ell_i$ to represent its realization.

Introduce the hard decision function $\theta(y)$ as $\theta(y) = 0$ if $y \geq 0$ and $1$ otherwise. For $\underline{Y}$, denote the vector of $[\theta(Y_1), \ldots, \theta(Y_n)] \in \mathbb{F}_2^n$ by $\theta(\underline{Y})$ for simplicity. As an exercise of Bayes' rule, we have
\begin{eqnarray*}
    &&\mbox{if\;} y_i \geq 0, P(W_i = 1 \mid Y_i = y_i) = \frac{1}{1 + \exp(\ell_i)}, \mbox{\;and}\\
    &&\mbox{if\;} y_i < 0, P(W_i = 0 \mid Y_i = y_i) = \frac{1}{1 + \exp(-\ell_i)}.
\end{eqnarray*}
So the conditional probability that a transmitted bit is flipped by the hard decision of the channel output is given by
\begin{equation}\label{eq:p_error}
    P(\theta(Y_i) \neq W_i \mid Y_i = y_i) = \frac{1}{1 + \exp(|\ell_i|)}.
\end{equation}

\subsection{GRAND}\label{2.2}

In a nutshell, the idea of GRAND is to find some $\underline{E} \in \mathbb{F}_{2}^{n}$, called an error pattern, so that by flipping the hard decisions $\theta(\underline{Y})$ according to $\underline{E}$, the result $\theta(\underline{Y}) \oplus \underline{E}$ is a codeword in $\mathcal{C}$.


In order to find such an error pattern that results in a codeword, a sequence of candidate error patterns are sequentially queried, until a codeword is reached, or until all candidate error patterns are exhausted. For ease of exposition, here we let the set of candidate error patterns be $\mathcal{E} = \{\underline{e}(1), \underline{e}(2), \ldots, \underline{e}(2^n)\} = \mathbb{F}_2^n$, in which each $\underline{e}(t)$ is a different vector in $\mathbb{F}_2^n$. Hence an error pattern $\underline{E}$ can surely be found, sooner or later. At the end of this subsection we will discuss how, in practice, we truncate the querying process due to complexity constraints.

The driving engine of a GRAND scheme is an ordering policy $\pi$, which is a permutation of $\{1, 2, \ldots, 2^n\}$, depending upon $\underline{y}$. A GRAND scheme thus sequentially checks whether the criterion $\theta(\underline{y}) \oplus \underline{e}(\pi(t)) \in \mathcal{C}$ is satisfied, for $t$ from $1$ to $2^n$. It stops as soon as it finds the first $t$ satisfying the criterion and declares the found codeword as the decoding result.

We generally determine the ordering policy $\pi$ using a reliability vector $\gamma(\underline{y}) = [\gamma_1(\underline{y}), \ldots, \gamma_n(\underline{y})]$, in such a way that, if any two candidate error patterns, say $\underline{e}(t)$ and $\underline{e}(t')$, satisfy
\begin{eqnarray}
    \sum_{i = 1, \ldots, n; e_i(t) = 1} \gamma_i(\underline{y}) \leq \sum_{i = 1, \ldots, n; e_i(t') = 1} \gamma_i(\underline{y}), 
\end{eqnarray}
then $\pi(t) \leq \pi(t')$. That is, candidate error patterns are sorted in ascending order in terms of their accumulated reliability values.

The description using the reliability metric $\gamma$ in the previous paragraph has led to a unified treatment of GRAND schemes \cite{liu2022orbgrand}. Different choices of $\gamma$ correspond to different GRAND schemes; for example, the original GRAND \cite{duffy2019capacity} has $\gamma_i(\underline{y}) = 1$, SGRAND \cite{solomon2020soft} (which is equivalent to ML decoding) has $\gamma_i(\underline{y}) = |\ell_i|$ (see also \cite[Ch. 10]{Lin2004ErrorCC}), ORBGRAND \cite{duffy2022ordered} has $\gamma_i(\underline{y}) = r_i$ where $r_i$ is the rank of $|\ell_i|$ among $\{|\ell_1|, \ldots, |\ell_n|\}$, in ascending order, and so on.
%

We say that an ordering policy $\pi$ is of ORB-type, if it only depends upon $\underline{r} = [r_1, \ldots, r_n]$, instead of the exact values of $\underline{y}$. ORB-type ordering policies inherit the key advantage of ORBGRAND, namely its amenability to hardware implementation: once the relationship of ranking among the magnitudes of LLRs is obtained and the channel output symbols are rearranged accordingly, the queries can then be sequentially conducted using a pre-generated sequence, which can be efficiently implemented in a recursive fashion \cite{duffy2022ordered}\cite{condo2021high}. Our aim is to design a new ORB-type ordering policy so as to approach the performance of SGRAND, which, unfortunately, is not of ORB-type.

\section{Search Problem and Reshuffled Querying}

In order to conduct a tractable analysis of the querying process in GRAND schemes, we turn to a search problem. Compared with decoding, the key simplification is that the search problem ignores the case where a querying process stops with a codeword different from the actually transmitted one. Our analysis for such an idealized situation sheds key insight into how an ordering policy should be designed.


\subsection{Search Problem Formulation}\label{Question setting}

We formulate the search problem as follows:

\begin{itemize}
\item For the channel model described in Section \ref{2.1}, let a codeword $\underline{W}$ be randomly chosen, so as to produce the BPSK transmitted vector $\underline{X}$ and the noise-corrupted channel output vector $\underline{Y}$, which is observed by the searcher.
\item The searcher adopts an ordering policy $\pi$, which is a permutation of $\{1, 2, \ldots, 2^n\}$, depending upon $\underline{Y}$.
\item The searcher sequentially conducts queries by computing $\theta(\underline{Y}) \oplus \underline{e}(\pi(t))$, for $t$ from $1$ to $2^n$. If for some $t$, $\theta(\underline{Y}) \oplus \underline{e}(\pi(t)) = \underline{W}$ is satisfied, a genie immediately informs the searcher that the chosen codeword $\underline{W}$ is found; otherwise, the searcher turns to the next $t$.

Here lies the key difference between the search problem and GRAND: the searcher does not care about whether any other codeword is encountered; --- instead, the searcher only cares about encountering $\underline{W}$ during querying.
\item For this idealized search problem, it is clear that there exists exactly one value of $t$ such that $\theta(\underline{Y}) \oplus \underline{e}(\pi(t)) = \underline{W}$ is satisfied, and the searcher stops once this $t$ is found.
\end{itemize}

\subsection{Analysis of Search Problem}

For the search problem, we are interested in the expected number of queries before the searcher stops. We emphasize that the expectation is with respect to the joint probability distribution of $(\underline{W}, \underline{Y})$, and hence it is an ensemble average result, given as follows.

\begin{theorem}\label{thm:Q}
    For the search problem in Section \ref{Question setting}, the expected number of queries before the searcher stops is given by
    \begin{equation}
        Q = \sum_{t=1}^{2^n} t \mathbf{E}_{\underline{Y}} [S_t], \label{eq:guessnum}
    \end{equation}
    where conditioned upon $\underline{Y} = \underline{y}$, $S_t$ is realized as
    \begin{align}
        s_t = P_{\underline{W}|\underline{Y}}\left(\theta(\underline{Y}) \oplus \underline{e}(\pi(t)) | \underline{Y} = \underline{y}\right),\label{eq:ML}
    \end{align}
    in which the subscript $\underline{W}|\underline{Y}$ emphasizes that $s_t$ is the posterior probability that the codeword is $\theta(\underline{Y}) \oplus \underline{e}(\pi(t))$ when the searcher observes $\underline{Y}$.
\end{theorem}
\begin{IEEEproof} According to the search problem described in Section \ref{Question setting}, the expected number of queries $Q$ can be calculated as

\begin{footnotesize}
    \begin{align*}
        & \sum_{\underline{w} \in \mathcal{C}} P(\underline{W} = \underline{w}) \int_{\mathbb{R}^n} p_{\underline{Y}|\underline{W}}(\underline{y}|\underline{w}) \cdot \sum_{t=1}^{2^n} t \cdot \mathbf{1}\left(\theta(\underline{y})\oplus \underline{w} = \underline{e}(\pi(t))\right) \text{d}\underline{y} \\
        & = \int_{\mathbb{R}^n} p(\underline{Y} = \underline{y}) \sum_{\underline{w} \in \mathcal{C}} P_{\underline{W}|\underline{Y}}(\underline{w} | \underline{y})  \sum_{t=1}^{2^n} t \cdot \mathbf{1}\left(\theta(\underline{y})\oplus \underline{w} = \underline{e}(\pi(t))\right) \text{d}\underline{y} \\
        & = \int_{\mathbb{R}^n} p(\underline{Y} = \underline{y}) \sum_{t=1}^{2^n} t \cdot P_{\underline{W}|\underline{Y}}\left(\theta(\underline{y}) \oplus \underline{e}(\pi(t)) | \underline{y}\right) \text{d}\underline{y} \\
        & = \sum_{t=1}^{2^n} t \cdot \int_{\mathbb{R}^n} p(\underline{Y} = \underline{y}) P_{\underline{W}|\underline{Y}}\left(\theta(\underline{y}) \oplus \underline{e}(\pi(t)) | \underline{y}\right) \text{d}\underline{y} \\
        & = \sum_{t=1}^{2^n} t \mathbf{E}_{\underline{Y}} [S_t] ,
    \end{align*}
\end{footnotesize}

\noindent where the second equality is due to that for given $\underline{y}$ and $\underline{e}(\pi(t))$, there is exactly one $\underline{w}$, namely $\theta(\underline{y}) \oplus \underline{e}(\pi(t))$, for which the indicator function is one.
%
\end{IEEEproof}

According to Theorem \ref{thm:Q}, the key to assessing a search problem is $S_t$, $t = 1, \ldots, 2^n$. When the elements of $\underline{W}$, $[W_1, \ldots, W_n]$, are independent and identically distributed (i.i.d.), $S_t$ can be evaluated using the following result.
\begin{lemma}\label{lemma-pt}
    When $[W_1, \ldots, W_n]$ are i.i.d., given $\underline{y}$, the posterior probability $s_t$ in \eqref{eq:ML} is given by
    \begin{eqnarray}
        s_t = \prod_{i:e_i(\pi(t)) = 1} \frac{1}{1+\exp(|\ell_i|)} \prod_{i:e_i(\pi(t)) = 0} \frac{\exp(|\ell_i|)}{1+\exp(|\ell_i|)}.\label{eq:p_i}
    \end{eqnarray}
\end{lemma}
\begin{IEEEproof} Due to the i.i.d. property of $\underline{W}$ and the memoryless property of the AWGN channel, an exercise of Bayes' rule yields
\begin{eqnarray}
    s_t &=& P_{\underline{W}|\underline{Y}}\left(\theta(\underline{Y}) \oplus \underline{e}(\pi(t)) | \underline{Y} = \underline{y}\right) \nonumber\\
    &=& \prod_{i = 1}^n P_{W_i|Y_i}\left(\theta(Y_i) \oplus e_i(\pi(t)) | Y_i = y_i\right).
\end{eqnarray}
Inspect the product. For those factors whose subscripts $i$'s satisfy $e_i(\pi(t)) = 1$, $\theta(Y_i)$'s are flipped, and according to \eqref{eq:p_error}, we have their product as $\prod_{i:e_i(\pi(t)) = 1} \frac{1}{1+\exp(|\ell_i|)}$; on the other hand, for those factors whose subscripts $i$'s satisfy $e_i(\pi(t)) = 0$, $\theta(Y_i)$'s are not flipped, and according to \eqref{eq:p_error}, we have their product as $\prod_{i:e_i(\pi(t)) = 0} \frac{\exp(|\ell_i|)}{1+\exp(|\ell_i|)}$. Multiplying these two parts together leads to \eqref{eq:p_i}.
\end{IEEEproof}

From the expression of $Q$ in \eqref{eq:guessnum}, we have that ordering policies minimizing the expected number of queries should render the sequence of $\{\mathbf{E}_{\underline{Y}} [S_t]\}_{t = 1, \ldots, 2^n}$ monotonically non-increasing.
\begin{corollary}\label{corollary-monotonic-pt}
    For the search problem, ordering policies that minimize $Q$ satisfy $\mathbf{E}_{\underline{Y}} [S_1] \geq \mathbf{E}_{\underline{Y}} [S_2] \geq \ldots \geq \mathbf{E}_{\underline{Y}} [S_{2^n}]$.
\end{corollary}
\begin{IEEEproof} This immediately follows from the expression \eqref{eq:guessnum}. \end{IEEEproof}

The expectation sequence $\{\mathbf{E}_{\underline{Y}} [S_t]\}_{t = 1, \ldots, 2^n}$ being monotone is a condition weaker than the sequence $\{s_t\}_{t = 1, \ldots, 2^n}$ itself being monotone for each $\underline{y}$. In fact, this latter condition corresponds to the ordering policy of SGRAND, as shown by the following result.

\begin{proposition}\label{Th:1}
The ordering policy $\pi$ that for each $\underline{y}$ rearranges queries so as to render $s_1 \geq s_2 \geq \ldots \geq s_{2^n}$ is identical to the ordering policy of SGRAND generated using $\gamma_i(\underline{y}) = |\ell_i|$ in Section \ref{2.2}.
\end{proposition}

\begin{IEEEproof} Consider any two candidate error patterns, $\underline{e}(t)$ and $\underline{e}(t')$, satisfying $s_t \geq s_{t'}$. Applying Lemma \ref{lemma-pt}, we have
\begin{eqnarray}
    && \prod_{i:e_i(\pi(t)) = 1} \frac{1}{1+\exp(|\ell_i|)} \prod_{i:e_i(\pi(t)) = 0} \frac{\exp(|\ell_i|)}{1+\exp(|\ell_i|)}\nonumber\\
    &\geq& \prod_{i:e_i(\pi(t')) = 1} \frac{1}{1+\exp(|\ell_i|)} \prod_{i:e_i(\pi(t')) = 0} \frac{\exp(|\ell_i|)}{1+\exp(|\ell_i|)}.\nonumber
\end{eqnarray}
Noting that the denominators on both sides of the inequality are in fact identical, i.e., $\prod_{i = 1}^n (1 + \exp(|\ell_i|))$, we cancel them and obtain
\begin{eqnarray}
    \prod_{i:e_i(\pi(t)) = 0} \exp(|\ell_i|) &\geq& \prod_{i:e_i(\pi(t')) = 0} \exp(|\ell_i|),\nonumber\\
    \mbox{i.e.,}\; \sum_{i:e_i(\pi(t)) = 0} |\ell_i| &\geq& \sum_{i:e_i(\pi(t')) = 0} |\ell_i|,\nonumber\\
    \mbox{i.e.,}\; \sum_{i:e_i(\pi(t)) = 1} |\ell_i| &\leq& \sum_{i:e_i(\pi(t')) = 1} |\ell_i|,
\end{eqnarray}
which is exactly the condition of $\pi(t) \leq \pi(t')$ in SGRAND, according to the description in Section \ref{2.2}.
\end{IEEEproof}



\subsection{Reshuffled Querying}\label{3.2}

Motivated by Proposition \ref{Th:1} and Corollary \ref{corollary-monotonic-pt}, we desire an ORB-type ordering policy that approximates the ordering policy of SGRAND in the sense that, although it is unlikely to have $\{s_t\}_{t = 1, \ldots, 2^n}$ be monotonically non-increasing for each $\underline{y}$, the overall effect, i.e., the ensemble average sequence $\{\mathbf{E}_{\underline{Y}}[S_t]\}_{t = 1, \ldots, 2^n}$ is monotonically non-increasing. So we propose the following simple approach:

\begin{itemize}
    \item We begin with an existing ORB-type ordering policy, denoted $\pi_\mathrm{base}$, for example, the original ORBGRAND \cite{duffy2022ordered} or CDF-ORBGRAND \cite{duffy2022ordered}\cite{liu2022orbgrand}, to obtain its corresponding expectation sequence $\{\mathbf{E}_{\underline{Y}} [S_t]\}_{t = 1, \ldots, 2^n}$.
    \item We then sort $\{\mathbf{E}_{\underline{Y}} [S_t]\}_{t = 1, \ldots, 2^n}$ so that they are in descending order. Denote the permutation achieving this goal by $\tilde{\pi}$.
    \item We then use the concatenation of $\pi_\mathrm{base}$ and $\tilde{\pi}$, $\pi = \tilde{\pi} \circ \pi_\mathrm{base}$, as the ordering policy for decoding.
\end{itemize}

We call the thus obtained decoder RS-ORBGRAND, where the prefix ``RS'' implies that it is obtained by reshuffling the ordering policy of a base ORBGRAND scheme. Since both $\pi_\mathrm{base}$ and $\tilde{\pi}$ are of ORB-type, RS-ORBGRAND is of ORB-type. In particular, we emphasize that, conditioned upon $\pi_\mathrm{base}$, $\tilde{\pi}$ is in fact a fixed permutation, even independent of the ranking $\underline{r}$, let alone $\underline{y}$. For obtaining $\tilde{\pi}$, only the expectation sequence $\{\mathbf{E}_{\underline{Y}} [S_t]\}_{t = 1, \ldots, 2^n}$ is needed, which can be evaluated offline, without requiring the exact realizations $\underline{y}$.

RS-ORBGRAND uses the reshuffling step $\tilde{\pi}$ to render the queries to satisfy Corollary \ref{corollary-monotonic-pt}. For each particular $\underline{y}$, this reshuffling may not always render the queries to satisfy $s_1 \geq s_2 \geq \ldots \geq s_{2^n}$. So there still exists a gap between RS-ORBGRAND and SGRAND. The following result sheds some further insight into this gap.

\begin{proposition}\label{Th:ExtraGuess1}
    For any ordering policy $\pi$, the expectation of its excess number of queries compared with SGRAND is given by
    \begin{equation}\label{eq:ExtraGuess2}
        \Delta Q = \mathbf{E}_{\underline{Y}}\left[\sum_{1\leq i<j\leq 2^n:S_i < S_j} (S_j - S_i) \right].
    \end{equation}
\end{proposition}
\begin{IEEEproof} Denote by $\hat{S}_t$, $t = 1, \ldots, 2^n$, the random variables characterized by \eqref{eq:ML} induced by SGRAND. In fact, $\{\hat{S}_t\}_{t = 1, \ldots, 2^n}$ is simply a rearrangement of $\{S_t\}_{t = 1, \ldots, 2^n}$ satisfying $\hat{S}_1 \geq \ldots \geq \hat{S}_{2^n}$. Applying the expression of $Q$ in Theorem \ref{thm:Q}, we have that the difference between the expected numbers of queries with the considered $\pi$ and SGRAND, i.e., the expectation of the excess number of queries, is given by
\begin{equation}\label{eq:ExtraGuess1}
    \Delta Q = \mathbf{E}_{\underline{Y}}\left[\sum_{t=1}^{2^n} t (S_t-\hat{S}_t)\right].
\end{equation}

To prove the equivalence between \eqref{eq:ExtraGuess2} and \eqref{eq:ExtraGuess1}, we compare their coefficients of $S_t$. Denote by $\rho_t$ the position of $S_t$ in $\{\hat{S}_t\}_{t = 1, \ldots, 2^n}$. In \eqref{eq:ExtraGuess1}, the coefficient of $S_t$ is $t - \rho_t$. On the other hand, in \eqref{eq:ExtraGuess2}, the coefficient of $S_t$ can be rewritten as $\sum_{i:i<t}\mathbf{1}(S_{i} < S_{t}) - \sum_{j:j>t}\mathbf{1}(S_{j} > S_{t})$.


We now take a closer look at how we rearrange $\{S_t\}_{t = 1, \ldots, 2^n}$ to obtain $\{\hat{S}_t\}_{t = 1, \ldots, 2^n}$. Starting with $S_t$, since there are $\sum_{j:j>t}\mathbf{1}(S_{j} > S_{t})$ elements with indices larger than $t$ and values also larger than $S_t$, we need to move them to ``the left of'' $S_t$, and since there are $\sum_{i:i<t}\mathbf{1}(S_{i} < S_{t})$ elements with indices smaller than $t$ and values also smaller than $S_t$, we need to move them to ``the right of'' $S_t$. This way, $S_t$ is moved to the position $\rho_t$ as $\hat{S}_{\rho_t}$. Therefore, we have
\begin{eqnarray*}
    && t + \sum_{j:j>t}\mathbf{1}(S_{j} > S_{t}) - \sum_{i:i<t}\mathbf{1}(S_{i} < S_{t}) = \rho_t,\\
    \mbox{i.e.,}\; && t - \rho_t = \sum_{i:i<t}\mathbf{1}(S_{i} < S_{t}) - \sum_{j:j>t}\mathbf{1}(S_{j} > S_{t}),
\end{eqnarray*}
meaning that the coefficients of $S_t$ in \eqref{eq:ExtraGuess2} and \eqref{eq:ExtraGuess1} are identical, for each $t = 1, \ldots, 2^n$.
\end{IEEEproof}

\begin{figure}[ht]
    \centering
    \includegraphics[width=0.48\textwidth]{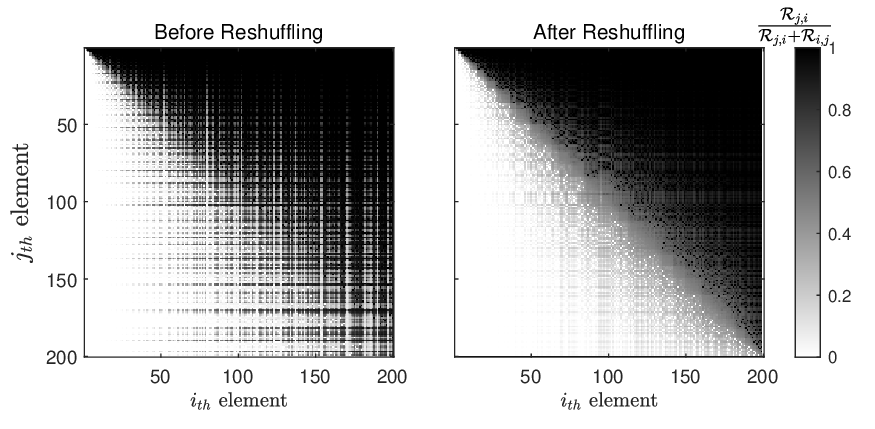}
    \caption{Illustration of $\mathcal{R}$ before and after applying the reshuffling step.}
    \label{fig:Gray}
\end{figure}

We can rewrite $\Delta Q$ in \eqref{eq:ExtraGuess2} as
\begin{eqnarray}
    \Delta Q = \sum_{i = 1}^{2^n} \sum_{j = i + 1}^{2^n} \mathbf{E}_{\underline{Y}} [(S_j - S_i) \cdot \mathbf{1}(S_j > S_i)],
\end{eqnarray}
and construct a $2^n \times 2^n$ matrix $\mathcal{R}$ with its $(j, i)$-th element as:
\begin{gather}
    \mathcal{R}_{j, i}=
	\begin{cases}
    \mathbf{E}_{\underline{Y}} \left[(S_j - S_i) \cdot \mathbf{1}(S_j > S_i)\right] \quad &i<j\\
	0 \quad &i=j\\
	\mathbf{E}_{\underline{Y}} \left[(S_i - S_j) \cdot \mathbf{1}(S_i > S_j)\right] \quad &i>j\\
	\end{cases}.\label{eq:m_ij}
\end{gather}
We see that $\Delta Q$ is the sum of the lower-triangular elements of $\mathcal{R}$. The reshuffling step $\tilde{\pi}$ in RS-ORBGRAND basically keeps swapping rows and columns of $\mathcal{R}$, so that if for any $i < j$, $\mathcal{R}_{j, i} > \mathcal{R}_{i, j}$, these two elements are interchanged. Using $\mathcal{R}$, we can visually illustrate the effect of the reshuffling step $\tilde{\pi}$. In Figure \ref{fig:Gray}, we display the first $200 \times 200$ rows and columns of $\mathcal{R}$, before and after applying the reshuffling step $\tilde{\pi}$ to the base $\pi_\mathrm{base}$, for which we choose to use CDF-ORBGRAND. Here for better visual effect, we plot in gray scale the normalized values $\mathcal{R}_{j, i}/(\mathcal{R}_{j, i} + \mathcal{R}_{i, j})$. It can be seen that after reshuffling, all the lower-triangular elements of $\mathcal{R}$ become no greater than 0.5, i.e., all the relatively large lower-triangular elements are interchanged with their upper-triangular counterparts.

In practice, since $2^n$ is typically an exceedingly large value, due to complexity constraints, we cannot afford the worst case. In implementation, we introduce two additional parameters, $T \leq T_1$ no greater than $2^n$. In RS-ORBGRAND, we apply the reshuffling step to only the first $T_1$ elements of $\{\mathbf{E}_{\underline{Y}}[S_t]\}_{t = 1, \ldots, 2^n}$. When decoding, we truncate the querying process if none of $\underline{e}(\pi(t))$, $t = 1, \ldots, T$, yields a codeword; --- when truncation occurs, no codeword is found and a decoding failure is declared.
As will be seen in the next section, choosing a value of $T_1$ sufficiently larger than $T$ is crucial for achieving a good performance. Noting that the reshuffling step is conducted offline, we can usually tolerate to use a sufficiently large value of $T_1$, given the abundant computing resources available to date.

\section{Performance Evaluation}

In this section, we apply RS-ORBGRAND proposed in the previous section to conduct some numerical studies. We consider two codes: BCH(127, 113) and polar(128, 114) with 10 cyclic redundancy check (CRC) bits. Unless specified otherwise, we use CDF-ORBGRAND \cite{duffy2022ordered} \cite{liu2022orbgrand} with $T_1 = 5 \times 10^4$ to generate the ordering policy, and when decoding, we permit at most $T = 10^4$ queries and terminate the querying process if no codeword is still not found by then.

\begin{figure}[htbp]
  \centering
  \includegraphics[width=0.48\textwidth]{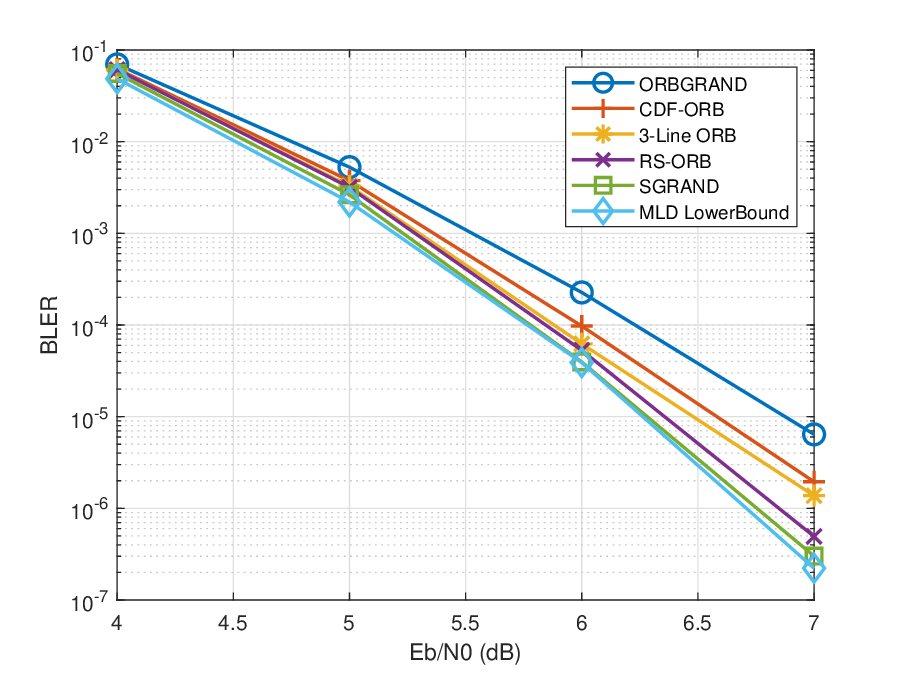}
  \caption{BLER for BCH(127, 113).}
  \label{fig:4.1}
\end{figure}

Figure \ref{fig:4.1} compares the BLER for BCH(127, 113) using several ORB-type decoders as well as SGRAND and ML decoding lower bound. The ML decoding lower bound is obtained by first running SGRAND with $T = 10^5$, and then supposing that even if no codeword has been found upon truncation, the decoding result would still be correct. It can be seen that only RS-ORBGRAND achieves performance close to SGRAND as SNR increases. When BLER is $10^{-6}$, the gain of RS-ORBGRAND compared with existing ORB-type decoders is at least 0.3dB, and the gap from the ML decoding lower bound is only 0.1dB. In Table \ref{Tab:2} we display the average number of queries for the decoders studied in Figure \ref{fig:4.1}, and we can observe that RS-ORBGRAND requires the least querying cost among ORB-type decoders. We remark that although SGRAND incurs the least average number of queries (as also suggested by the analysis in Section \ref{3.2}), its queries needs to be generated online based on the exact values of LLRs and hence its querying complexity is by far the highest.


%

\begin{table}[!t]
    \renewcommand{\arraystretch}{1.4}
    \caption{Average number of queries for BCH(127, 113).}
    \label{Tab:2}
    \centering
    \begin{tabular}{|c|c|c|c|c|}
    \hline
                  & 4dB      & 5dB     & 6dB    & 7dB    \\ \hline
    ORBGRAND\cite{duffy2022ordered}      & 790.8 & 83.89 & 7.072 & 1.479 \\ \hline
    CDF-ORBGRAND\cite{duffy2022ordered} & 727.9 & 67.44 & 5.476 & 1.478 \\ \hline
    3 Line-ORBGRAND\cite{duffy2022ordered} & 730.0 & 62.34 & 4.732 & 1.445 \\ \hline
    RS-ORBGRAND (proposed)  & 715.6 & 60.63 & 4.445 & 1.350 \\ \hline
    SGRAND\cite{duffy2021guessing}        & 666.5 & 52.99 & 3.932 & 1.328 \\ \hline
    \end{tabular}
\end{table}


\begin{figure}[!t]
    \centering
    \includegraphics[width=0.48\textwidth]{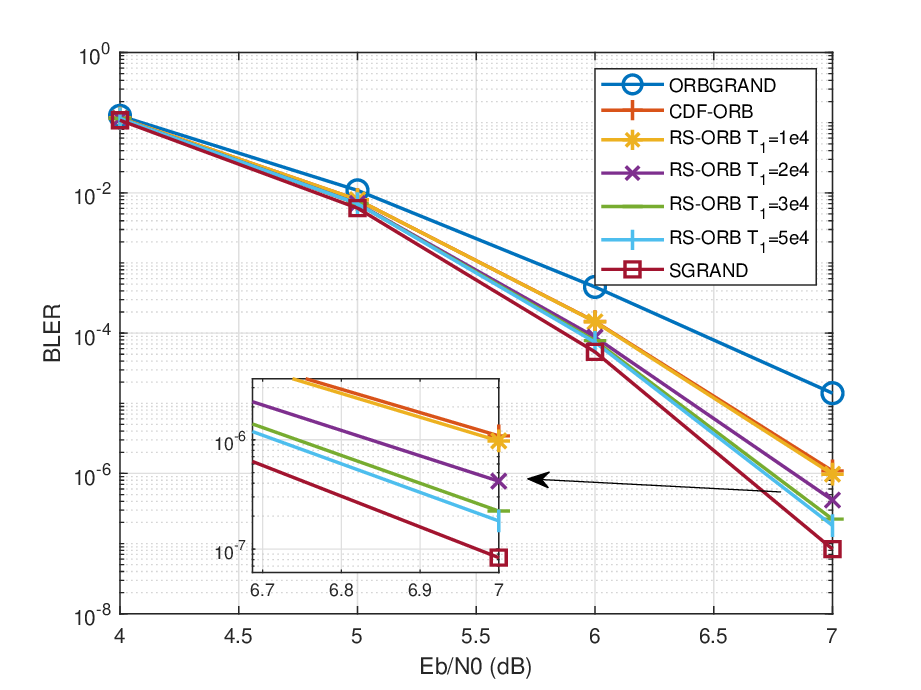}
    \caption{BLER for CRC-aided Polar(128, 114).}
    \label{fig:4.3}
\end{figure}

Figure \ref{fig:4.3} compares the BLER for CRC-aided polar(128, 114) using different decoders. Here for RS-ORBGRAND, we further study several different values of $T_1$, from $5 \times 10^4$ down to $10^4$. It is clearly shown that using a sufficiently large set of candidate error patterns is crucial. Again, we note that $T_1$ candidate error patterns are used offline for obtaining the reshuffling step, but when decoding, all the decoders can conduct at most $T = 10^4$ queries. By using a large value of $T_1$, we can identify more relatively large $\mathbf{E}_{\underline{Y}} [S_t]$'s and reshuffle them, so as to have them queried earlier in RS-ORBGRAND.

%
%

\section{Conclusion}

RS-ORBGRAND, as an improvement of ORBGRAND, is proposed and studied. Its basic idea is to reshuffle queries with the goal of approximating SGRAND. This improved scheme still inherits the key advantage of ORBGRAND, namely its dependency upon the ranking of the magnitudes of LLRs only. Numerical simulations show that RS-ORBGRAND achieves noticeable gain compared with existing ORB-type decoders, and is very close to ML decoder even in the regime of low BLER.



\newpage
\bibliographystyle{IEEEtran}
\bibliography{Ref}

\newpage
\clearpage
\appendices
{
}

\end{document}